# Anisotropic Curie temperature materials

Jason N. Armstrong, Susan Z. Hua, and Harsh Deep Chopra*

*Laboratory for Quantum Devices, Materials Program, State University of New York at Buffalo, NY 14260, USA*

Existence of anisotropic Curie temperature materials [E. R. Callen, *Phys. Rev.* **124**, 1373 (1961)] is a longstanding prediction – materials that become paramagnetic along certain crystal directions at a lower temperature while remaining magnetically ordered in other directions up to a higher temperature. Validating Callen's theory, we show that all directions within the basal plane of monoclinic $Fe_7S_8$ single crystals remain ordered up to 603 K while the hard c-axis becomes paramagnetic at 225 K. Materials with such a large directional dependence of Curie temperature opens the possibility of uniquely new devices and phenomena.

Closely related to magnetocrystalline anisotropy[1] is the anisotropy of magnetization.[2, 3] The latter's origins lie in magnetic anisotropy promoting the alignment of spins within a narrow cone along the easy axis while spreading them (wider cone) along the hard direction. This leads to a smaller value of saturation magnetization along the hard axis compared to the easy axis, *cf*. Fig. S1 in supplementary document).[4] While this difference is negligible in ordinary ferromagnets (~0.1% in Co),[3] it can become appreciable in materials with large magnetic anisotropy and/or small exchange energy.[5] In formulating the theory of anisotropic magnetization, Callen and Callen predicted the existence of anisotropic Curie temperature.[3] Simply defined, along the easy axis, the magnetic anisotropy tends to hold the spin cone and raises the Curie temperature, while along the hard axis anisotropy tends to spread the spin cone and lowers the Curie temperature.[6, 7] Perturbation theory is inadequate in formulating this effect and Callen used the quantum mechanical internal field Hamiltonian to show that for large ratios of magnetocrystalline anisotropy energy to the exchange energy, the moment along the hard axis drops abruptly to zero at a temperature that is lower than the Curie temperature along the easy axis.[6, 7] In comparison, the anisotropy of Curie temperature is negligible in ordinary ferromagnets due to the vanishingly small ratio of anisotropy to exchange energy, i.e., the spherical or isotropic exchange term in the denominator dominates. Candidate materials for large anisotropy of Curie temperature would have low crystal symmetry (which promotes higher magnetocrystalline anisotropy) and/or have weak exchange. A promising candidate meeting these criteria is the ferrimagnetic compound $Fe_7S_8$ (pyrrhotite). It has a pseudohexagonal structure that is slightly monoclinic with 8 molecules of $Fe_7S_8$,[8] and its investigation was suggested by Callen.[7] Its crystal and magnetic structure is discussed in the supplementary document.[4] The experimental details used to analyze the existence of anisotropic Curie temperatures are also described in detail in the supplementary document.[4] In the following, the magnetic ordering within the basal plane of $Fe_7S_8$ is considered first followed by behavior along the c-axis.

**Basal plane characteristics:** The fundamental property of interest in characterizing a magnetic phase at a given temperature is its spontaneous magnetization, $\sigma_{o,T}$. The temperature dependence of $\sigma_{o,T}$ is shown in Fig. 1(a) using the magnetic isotherm method; see discussion on magnetic isotherms in supplementary document.[4] The inset in Fig. 1(a) shows a zoom-in view of the curve at low temperatures to highlight the transition at 34 K that is characteristic of the $Fe_7S_8$ composition.[9] The key point of Fig. 1(a) is that *the basal plane remains spontaneously magnetized at all temperatures below ~600 K*. Previously, Curie temperature of $Fe_7S_8$ has been determined by a variety of techniques,[10-13] with most studies finding a Curie temperature ~600 K; see also Ref. [10] by Lotgering for a detailed discussion on the role of sulfur in the observed variation in Curie temperature. In one study, a Curie temperature of 565 K and an anomalous susceptibility behavior above it was reported.[14] However, those results were subsequently shown by Lotgering to be a direct result of magnetite impurities ($Fe_3O_4$) in $Fe_7S_8$.[10] We measured the Curie temperature of six samples from two different crystals by extrapolation of the steepest portion of the square of the spontaneous magnetization versus temperature ($\sigma_{o,T}^2$ versus $T$), Fig. 1(b), and by calorimetry, inset of Fig. 1(b). Figure 1(b) shows that the Curie temperature of the crystals is ~603-604 K, in close agreement with previous studies.

**Magnetic behavior along c-axis:** Figure 2(a-f) shows examples of magnetization curves along the c-axis at various temperatures. Higher temperature curves are characterized by linear magnetization curves, *cf.*, Fig. 2(a). Although shown for fields up to 7 T, this proportionality persist even when the samples were measured in a 9 T magnet. At low temperatures, a spontaneous magnetization gradually emerges and superimposes itself on the linear magnetization curve. The magnitude of this spontaneous magnetization increases as the temperature is lowered, and is accompanied by widening of the hysteresis loops, Fig. 2. This "new phenomenon" has previously been noted by Néel.[15] To investigate the appearance of spontaneous magnetization along the c-axis Arrott curves were plotted using the measured magnetization data along the c-axis. Briefly, Arrott curves are used to determine the Curie temperature in ferro- or ferrimagnets, by plotting $\sigma_{H,T}^2$ versus $H/\sigma_{H,T}$ isotherms,[16] as shown in Fig. 3. Care was taken to use fields that are sufficiently large to erase magnetic domains and hysteretic effects.[17, 18] The linear asymptote (red lines) of the isotherm that passes through the origin defines the Curie temperature. From Fig. 3, Arrott curves have a positive curvature above 190 K, and a negative curvature below 180 K, clearly revealing a transition from paramagnetism to ferrimagnetism in between. Extrapolation of the linear asymptotes yields a Curie temperature of 186 K. For temperatures higher than the Curie point, the intercept along abscissa ($H/\sigma_{H,T}$) in Fig. 3 gives the inverse susceptibility $1/\chi$; at the Curie temperature, $1/\chi \to 0$ ($\chi \to \infty$). For temperatures lower than the Curie point, the intercept along the y-ordinate gives square of spontaneous magnetization $\sigma_{o,T}^2$. Their temperature dependence is shown in the inset of Fig. 3. Whereas the Arrott curves clearly establish the existence of paramagnetic to ferrimagnetic transition along the c-axis, the exact value of Curie temperature is higher than 186 K because these isotherms are plotted as a function of applied field (as is the convention) without taking into account the demagnetization field. This along with other factors are well known to cause the Curie temperature to be underestimated.[16, 19] Instead of estimating the internal field of the sample and without a clear knowledge of exchange force at play in these materials at the present time, it is not only prudent but it should also be possible to independently corroborate the Curie temperature along the c-axis by direct measurement of other physical properties, and thus determine its precise value. This was done by measuring the temperature dependence of heat capacity $C_P$ using high sensitivity ac-calorimetery, Fig. 4(a). The heat capacity shows a sharp lambda transition at 225 K. Due to increasing fluctuations in magnetic alignment, $C_P$ increases as the temperature is raised, and then drops abruptly at the Curie point. The observed lambda transition also matches with peaks in the measured ac and dc susceptibility, as shown in Fig. 4(b-c), respectively. With hindsight, it is of interest to note that a small anomaly can be seen in the precise neutron diffraction data in Ref. [13] close to 225 K, but went unnoticed given a different focus of that study. On the other hand, the lambda peak (which is less than 1.5 K wide) is too narrow to be detected in the heat capacity measurements on $Fe_7S_8$ in Ref. [20] where data was taken every 10 K in this temperature range, and where the focus was on low temperature transition at 34 K.

In conventional crystals the Curie temperature is virtually independent of the crystal direction due to the miniscule ratio λ of the anisotropy energy to the isotropic exchange energy (~$10^{-5}$ for Fe and Ni, and ~$10^{-3}$-$10^{-4}$ in Co). The exchange energy for Fe and Co is ~700 $cm^{-1}$ and $10^3$ $cm^{-1}$, as compared to thermal energy $kT$ of 200 $cm^{-1}$ at room temperature. For $Fe_7S_8$, an estimate for λ can be made by plotting the measured value of spontaneous magnetization along the c-axis along with Callen's predicted values. Callen has shown that spontaneous magnetization $\langle S_\zeta \rangle$ as a function of temperature $(1/a)$ can be expressed by the following *implicit* function:[6, 7]

$$\langle S_\zeta \rangle = \frac{e^{a\langle S_\zeta \rangle} - e^{-a\langle S_\zeta \rangle}}{e^{a\langle S_\zeta \rangle} + e^{-a\langle S_\zeta \rangle} + e^{a\lambda}} \quad [1]$$

For $\lambda \leq 0.4621$, the Curie temperature is given by:

$$a_C = 1 + \tfrac{1}{2} e^{\lambda a_C} \quad [2]$$

Figure 5 plots magnetization along the c-axis for various values of λ (solid lines). The reduced coordinates for temperature $(1/a)$ from 0 to 1 scale with the ratio of thermal energy over exchange energy. In Fig. 5, negative values of λ correspond to easy axis along the c-axis, whereas positive values signify c-axis being the hard direction; the case of λ equal to zero refers to an isotropic materials. Equation [2] shows that for large negative values of λ, $a_c \to 1$. Therefore, $kT_c$ approaches the exchange energy; $k$ is the Boltzmann constant. As $\lambda \to 0$, $1/a_c \to \frac{2}{3}$, that is, $kT_c \to \frac{2}{3} \times \text{exchange energy}$. As the value of λ increases (becomes more positive), the anisotropy will cause the spin cone along the c-axis to spread and the magnetic order will disappear at a lower value of thermal energy, i.e., less $kT$ is needed to break the cone of magnetization, resulting in a lower Curie temperature along the c-axis. Figure 5 shows that the measured values of magnetization along the c-axis correspond to $\lambda$ greater than 0.462.

To conclude, unlike conventional magnetic materials, the anisotropy of Curie temperature raises intriguing and interesting questions. For example, the role of magnetic domain structure within the basal plane in influencing the observation of the lower Curie temperature remains to be investigated. From the viewpoint of electron transport, $Fe_7S_8$ may be regarded as an ultimate magnetic multilayer with each monolayer of iron plane separated by a monolayer of a sulfur plane, raising the possibility of interesting electron transport behavior.

**Acknowledgments:** This work was supported by the National Science Foundation, Grant Nos. DMR-0706074, and DMR-0964830, and this support is gratefully acknowledged.
*Corresponding author: H.D.C.; E-mail: hchopra@buffalo.edu

**FIGURE CAPTIONS**

**FIG. 1**. (a) Temperature dependence of spontaneous magnetization, as determined from basal plane measurements. Inset shows a zoom-in view of the low temperature transition at 34 K. (b) Determination of Curie temperature by extrapolation of the steepest portion of square of spontaneous magnetization versus temperature. The lower-left inset is a differential scanning calorimetry trace on heating the sample, showing Curie transition at 603 K.

**FIG. 2**. (a-f) Examples of specific magnetization versus field curves at various temperatures along the c-axis.

**FIG. 3**. Arrott curves ($\sigma_{H,T}^2$ versus $H/\sigma_{H,T}$) at various temperatures in the vicinity of Curie temperature, using the measured data along the c-axis. Inset shows spontaneous magnetization

$\sigma_o$ and inverse susceptibility versus temperature below and above the Curie temperature, respectively.

**FIG. 4**. (a) Temperature dependence of specific heat $C_P$ measured by ac-calorimetry. (b) ac susceptibility, measured at 2 T bias field, and 7 Oe, 1 kHz ac field. (c) dc susceptibility. The dc susceptibility represents the slope of the linear portion of the curves along the c-axis at high fields.

**FIG. 5**. Spontaneous magnetization $\langle S_{c-axis} \rangle$ versus temperature $(1/a)$ using Eq. 1 (solid lines). The measured values are shown by solid symbols.

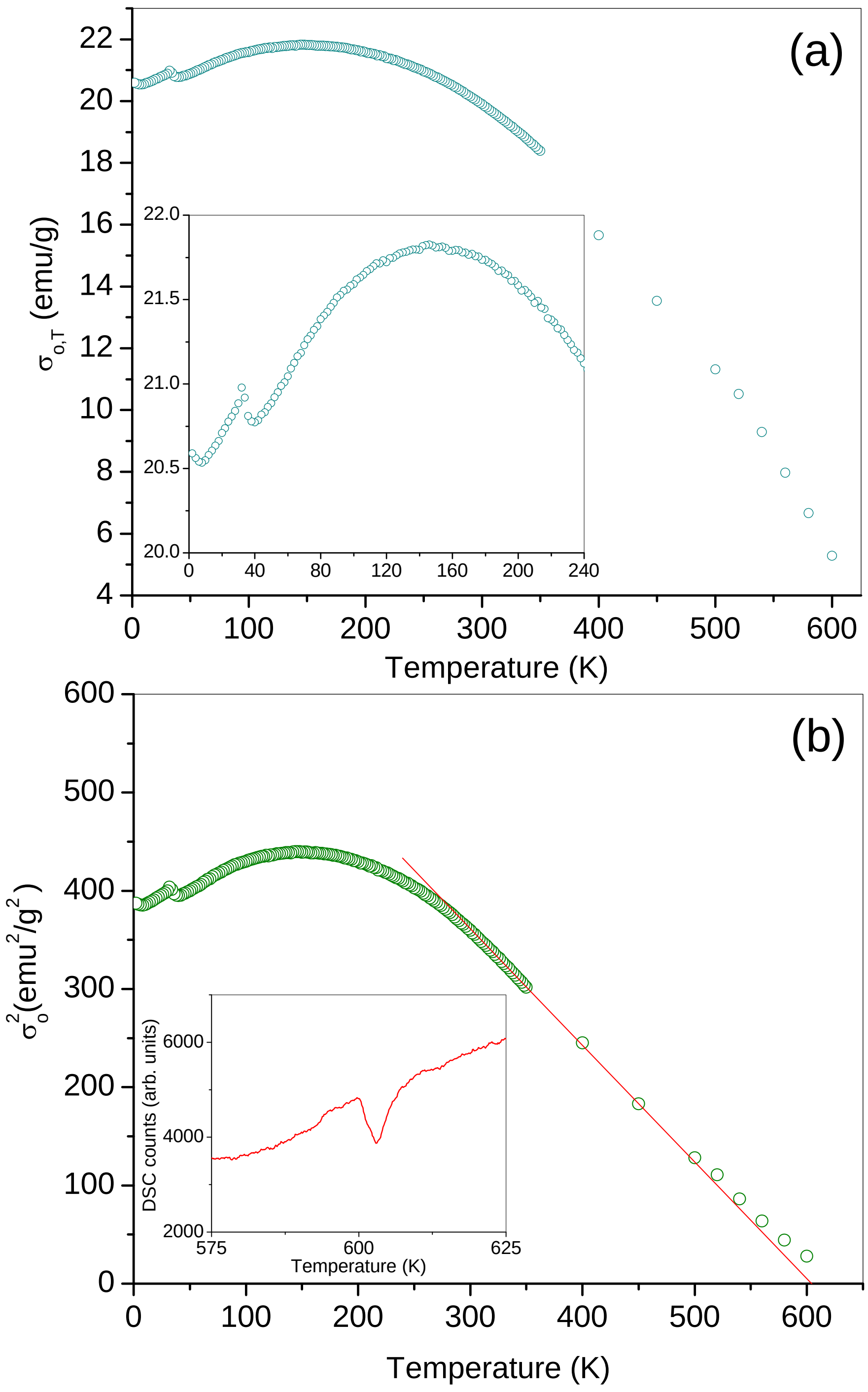

Figure 1

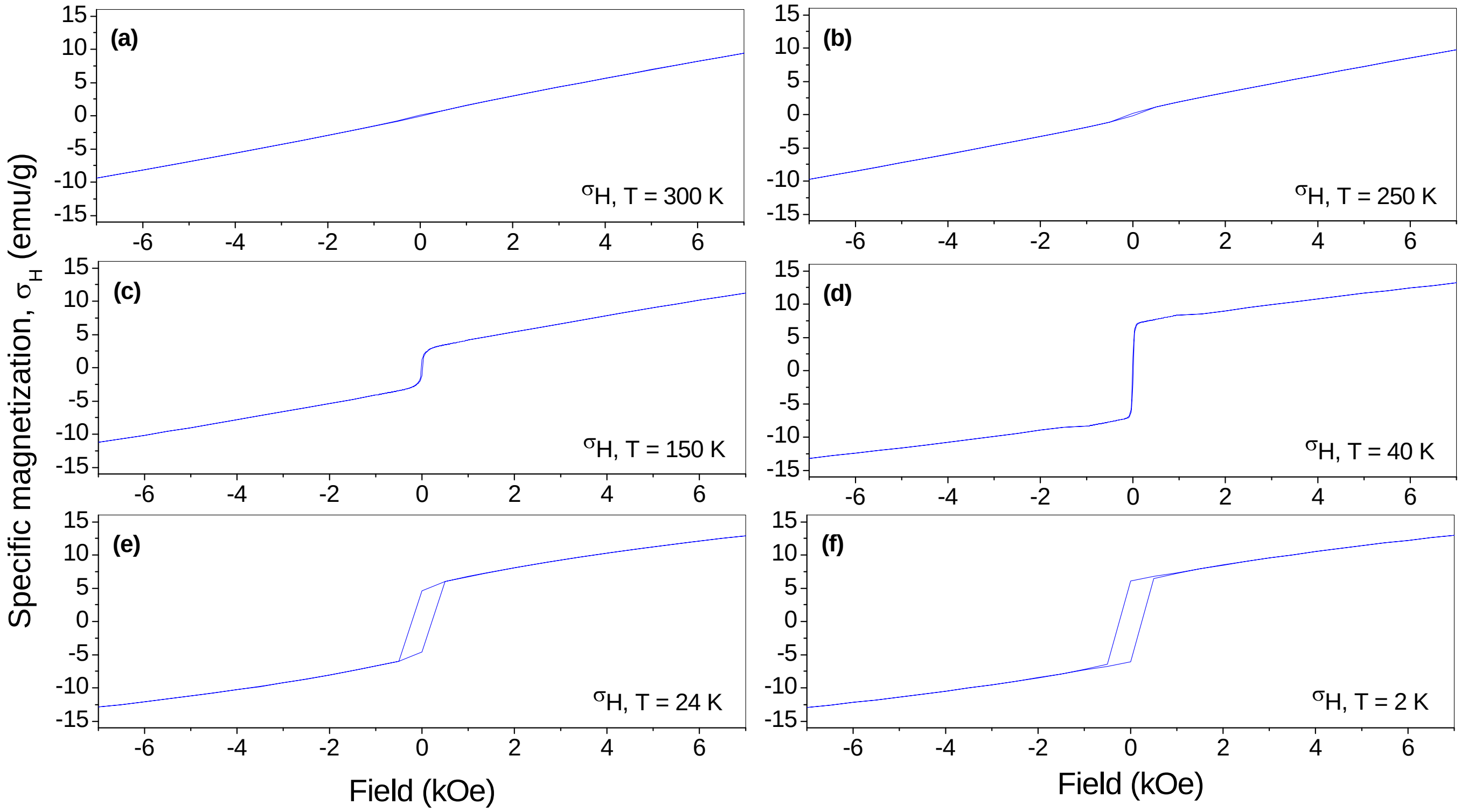
Specific magnetization, $\sigma_H$ (emu/g)
(a)
$\sigma$H, T = 300 K
(b)
$\sigma$H, T = 250 K
(c)
$\sigma$H, T = 150 K
(d)
$\sigma$H, T = 40 K
(e)
$\sigma$H, T = 24 K
(f)
$\sigma$H, T = 2 K
Field (kOe)
Field (kOe)

Figure 2

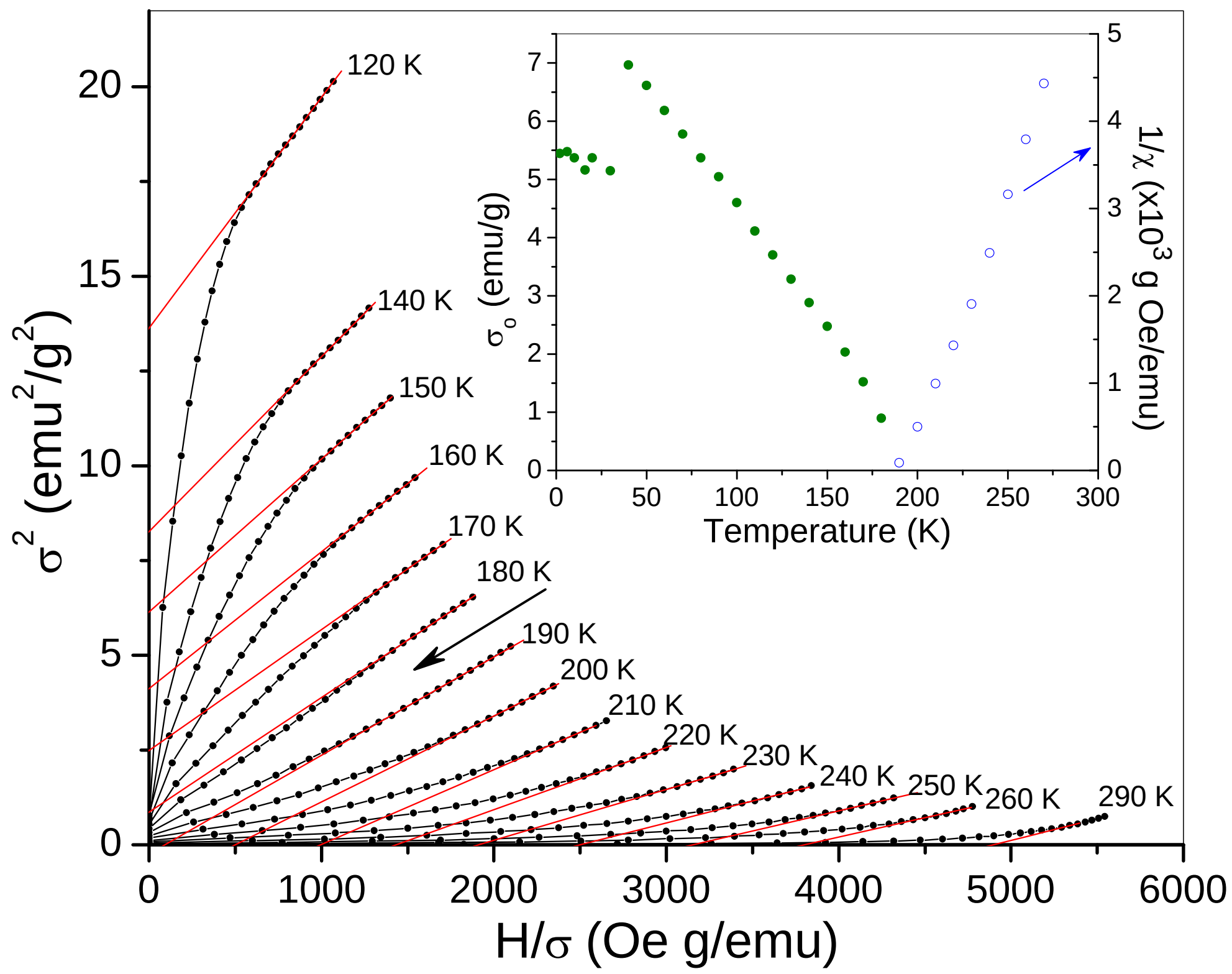

Figure 3

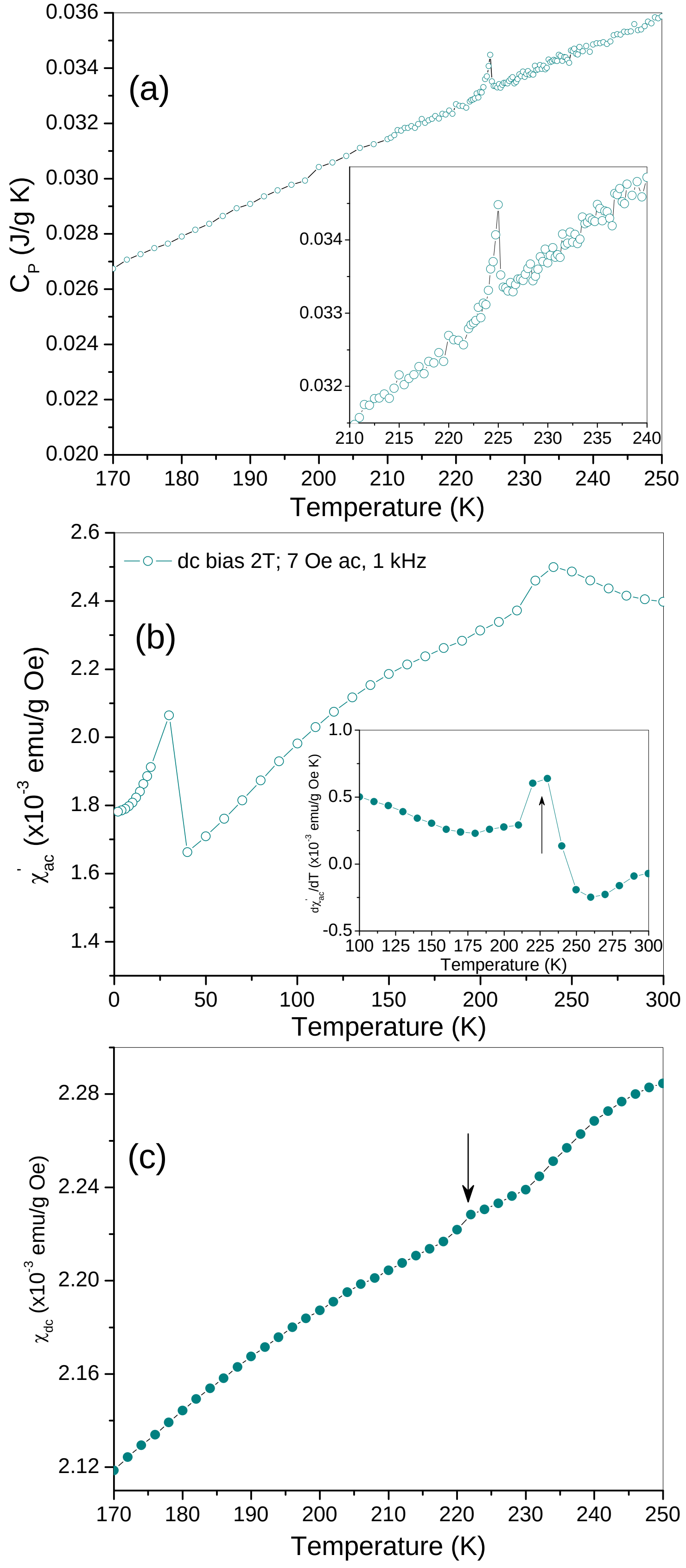

(a)
$C_P$ (J/g K)
Temperature (K)
(b)
dc bias 2T; 7 Oe ac, 1 kHz
$\chi'_{ac}$ (x10$^{-3}$ emu/g Oe)
$d\chi'_{ac}/dT$ (x10$^{-3}$ emu/g Oe K)
Temperature (K)
(c)
$\chi_{dc}$ (x10$^{-3}$ emu/g Oe)
Temperature (K)

Figure 4

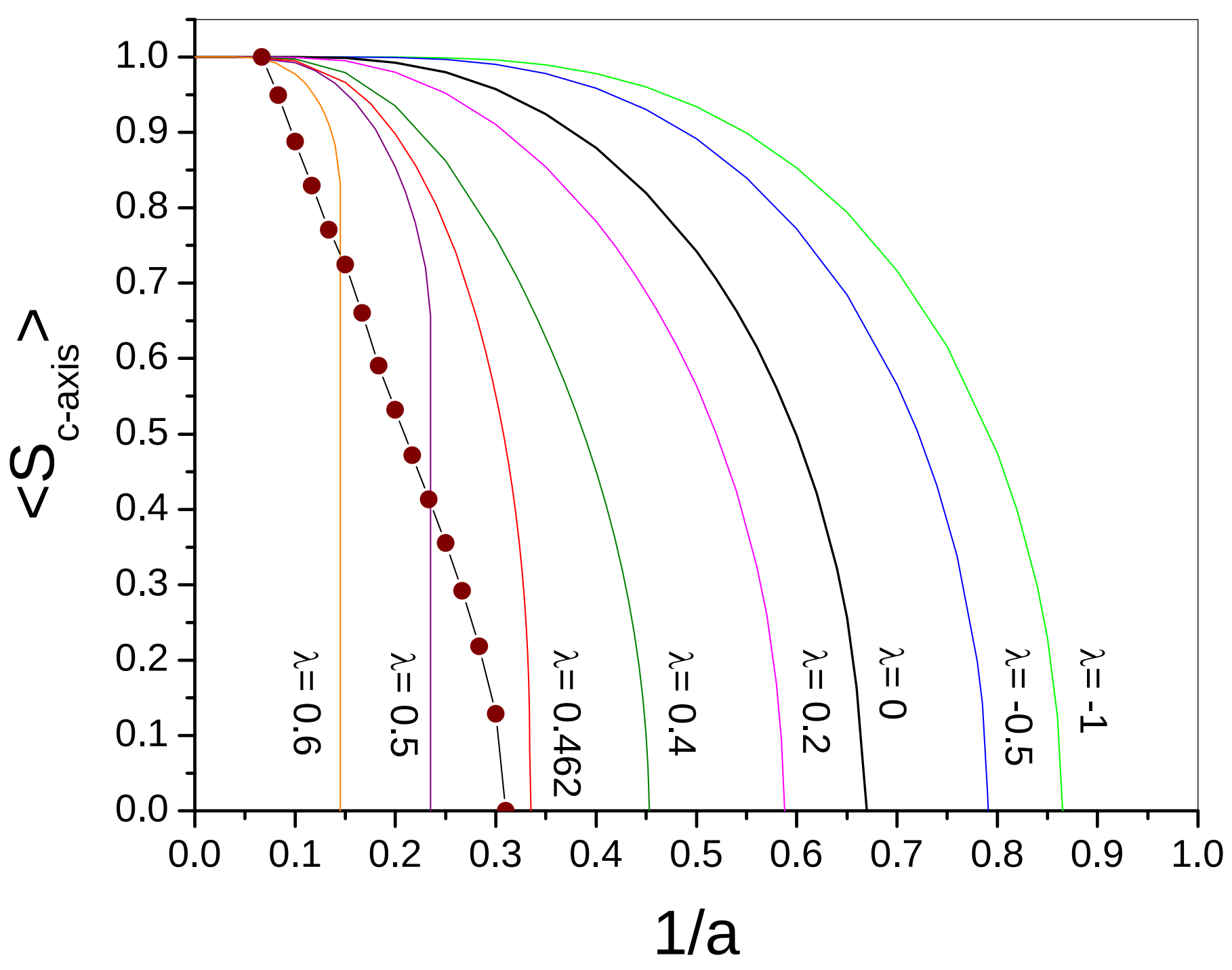

Figure 5

Anisotropic Curie temperature materials, Jason N. Armstrong, Susan Z. Hua, and Harsh Deep Chopra, Laboratory for Quantum Devices, Materials Program, State University of New York at Buffalo, NY 14260, USA

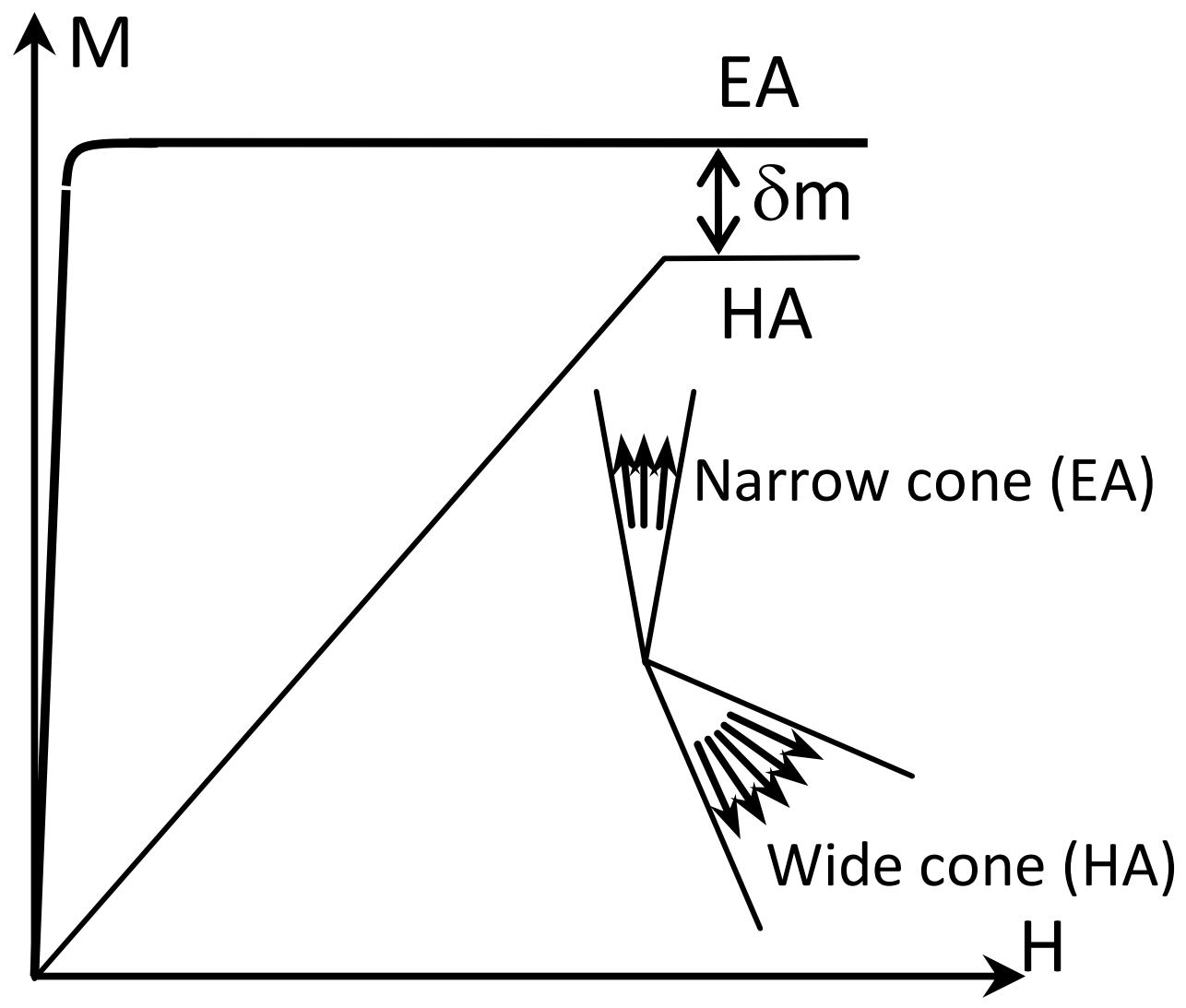

**Figure S1.** Schematic illustrating the anisotropy of magnetization. Note that saturation magnetization is lower along hard axis compared to easy axis. See main text for discussion.

**Crystal Structure of $Fe_7S_8$:** The crystal structure of $Fe_7S_8$ has been continually refined using x-ray, neutron diffraction, and TEM.[1-8] Bertaut investigated the structure of $Fe_7S_8$,[3, 9] and found the diffraction spots from $Fe_7S_8$ to be inconsistent with the NiAs unit. Instead it has a pseudohexagonal structure that is slightly monoclinic with 8 molecules of $Fe_7S_8$. Along the pseudohexagonal c-axis, the superstructure of $Fe_7S_8$ consists of iron planes separated by sulfur planes. With additional considerations of electrostatic exchange forces between iron atoms in iron-deficient $Fe_7S_8$, Néel concluded that this crystal is best represented as $Fe_7S_8L$, where L is a vacancy.[10] This gives rise to two sub-lattices for iron that are crystallographically different due to ordered vacancies in alternate iron layers normal to the c-axis. This asymmetry of iron sub-lattices also gives rise to ferrimagnetism and a magnetic symmetry within the basal plane that is lower than the crystal symmetry.[10] Neutron diffraction studies show that asymmetry of the two iron sub-lattices due to ordered vacancies gives rise to ferrimagnetism; the moments lie in the (001) planes, with anti-parallel orientation in adjacent Fe planes.[8, 11]

**Magnetic Structure of $Fe_7S_8$:** Weiss reported magnetization curves along the basal plane of $Fe_7S_8$, with not enough results along the c-axis of $Fe_7S_8$.[12] To extend Weiss' data, Pauthenet measured magnetization along the c-axis, and found a linear behavior in fields up to ~2 T at room temperature.[13] At lower temperatures, Pauthenet observed the appearance of a spontaneous magnetization along the c-axis. Néel described this "new phenomenon" as progressive superposition of a spontaneous magnetization on the "preceding paramagnetism" of the c-axis.[10] Magnetocrystalline anisotropy measurements show that the crystal exhibits a small triaxial symmetry in the basal plane,[14] and the in-plane anisotropy constants have recently been refined.[15] In contrast, only a small linear increase in moment is observed along the c-axis even in fields as high as 10 T.[16, 17] The unusual behavior has been linked to ordered vacancies in alternate iron layers normal to c-axis.[18] A two-sublattice rotation molecular field model has also been developed to explain the linear behavior of $Fe_7S_8$ along the c-axis.[19]

Note that although Weiss, Néel, and others frequently referred the c-axis of $Fe_7S_8$ as being 'paramagnetic' to describe its linear characteristics and to draw contrast with the ferrimagnetic basal plane,[10, 12-14, 19-21] their use of the term 'paramagnetic' does not imply the existence of a different magnetic ordering temperature along the c-axis compared to the rest of the crystal. It was Carr who initially suggested the possibility of materials with anisotropic Curie temperature (as discussed on page 320 of Ref. [22]), and Callen developed the theory of magnetic materials with anisotropic Cure temperature.[23, 24]

Anisotropic Curie temperature materials, Jason N. Armstrong, Susan Z. Hua, and Harsh Deep Chopra, Laboratory for Quantum Devices, Materials Program, State University of New York at Buffalo, NY 14260, USA

## Experimental Details

Measurements were made on six different samples carved from two different natural crystals of $Fe_7S_8$. The chemical composition was confirmed using EDAX, and it showed that the nominal composition of all the samples were close to $Fe_{0.875}S$ ($Fe_7S_8$ ), with composition varying slightly from $Fe_{6.86}S_8$ to $Fe_{7.26}S_8$. The magnetization curves from 2 K to 350 K were measured in a Quantum Design PPMS with a 7 T magnet. Another PPMS that was equipped with a 9 T magnet and high temperature VSM was used for measurements up to 640 K. A room temperature VSM with a maximum field of ~0.88 T was also used for preliminary analyses. The 7 T PPMS was used to measure the ac susceptibility, with bias field ranging up to 2 T, an ac field of 7 Oe, and excitation frequency from 10 Hz to 1 kHz. The magnetic moment and susceptibility was calibrated using a standard Pd sample. In $Fe_7S_8$, the c-axis is paramagnetic, i.e., magnetization is entirely confined in the basal plane. Therefore, measurement of magnetization along any arbitrary angle away from the c-axis simply measures the vector projection of basal plane magnetization along that direction. Therefore, it is of critical importance to precisely align the magnetic field along the c-axis. This is non-trivial and must be stringently adhered to in experiments. We performed numerous test runs to ensure precise alignments. To prevent rotation at high field, clamping is required.

The specific heat $C_P$ of $Fe_7S_8$ was measured by ac-calorimetry[25, 26] using a microfabricated chip, and the measurements were repeated 3 times in the temperature ranges of interest. As shown in Fig. S2, the ac calorimeter consists of a thin film Pt heater and a sensor, both 100 nm thick, deposited on a 200 nm thick silicon nitride membrane supported by a silicon frame.

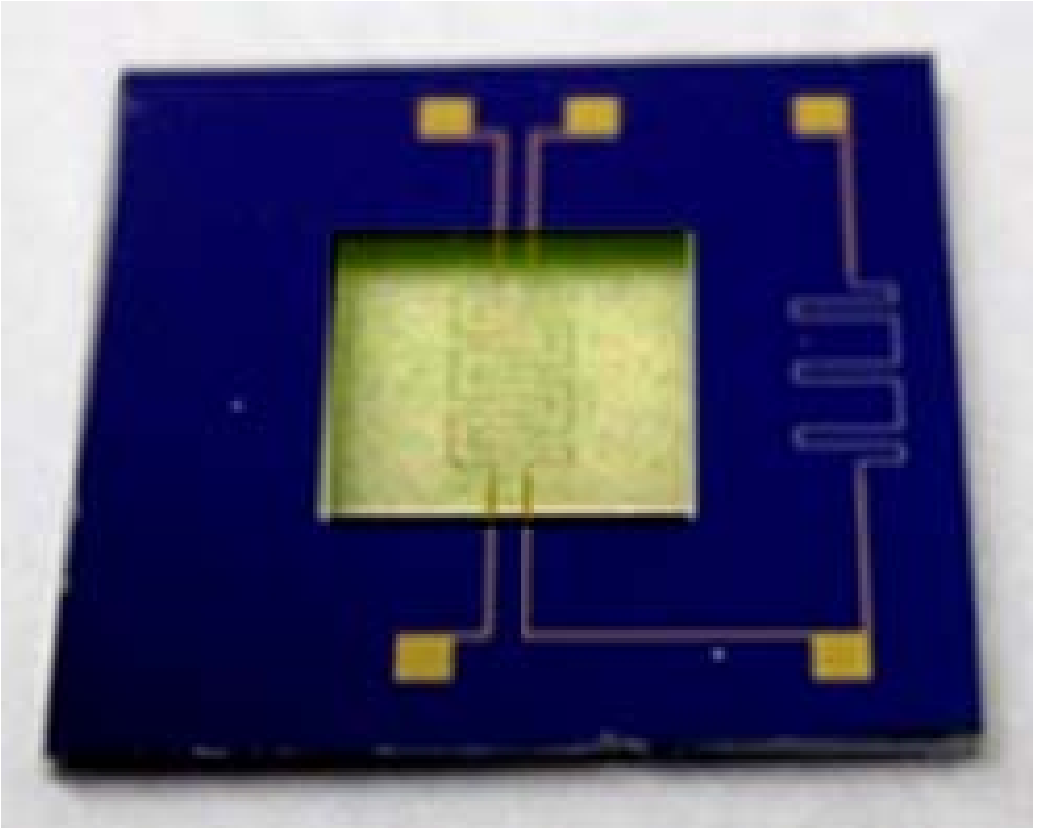

**Figure S2**. Microfabricated ac calorimetry chip for measurement of heat capacity.

The heater and sensor elements on the membrane were connected to gold leads on the silicon frame. A Keithely 6221 ac and dc current source was used for heating at a frequency of 2.25 Hz and amplitude of 0.4 mA. The temperature change was measured by passing a dc current (10 μA) through the sensor using a Keithley 6221 current source, and measuring the corresponding voltage change using a lock-in amplifier (Stanford Research System SR830). The sample was fixed on the ac chip using N-grease (Apiezon). Prior to the measurements, the heat capacity of the chip with N grease was also measured to determine the heat addenda and also to rule out any extraneous peaks as a function of temperature. This design of ac calorimeter is similar to the one described in Ref. [27]. The measurements were made in the PPMS under vacuum ($10^{-6}$ Torr). Use of vacuum instead of high pressure He (~10 mtorrs) in PPMS significantly reduces drift and heat loss to the environment. For the average transmitted power $P_o$ from the heater, the amplitude of temperature oscillations $T_{ac}$ is given by:[25]

$$T_{ac} = \frac{P_o}{\omega C}\left[1 + \frac{1}{(\omega\tau_1)^2} + (\omega\tau_2)^2 + \frac{2K_b}{3K_s}\right]^{-1/2} = \frac{P_o}{\omega C}F(\omega) \qquad [1]$$

The time constant $\tau_1$ characterizes the chip-to-bath relaxation time. The time constant $\tau_2$ characterizes the response time of the sample and addenda (heater, thermocouple, and membrane platform) to a heat input, and $C$ is the total capacity of the sample and addenda. When $\omega\tau_1 \gg$ $1 \gg \omega\tau_2$, and the thermal conductivity of the sample $K_s$ is kept large with respect to sample-to-bath thermal conductivity $K_b$, then $F(\omega) = 1$, and, Eq. 1 takes a simple form for the heat capacity of the sample $C_P \cong P_o/\omega T_{ac}$. In practice, the optimum operating frequency $\omega$ is determined by measuring the frequency dependence of $T_{ac}$ to obtain the 'adiabatic plateau' by plotting $\omega T_{ac}$ versus frequency within the range of inequality,[28] and the operating frequency ω is typically between $10^{-1}$-$10^{4}$ Hz depending on the chip setup and sample.[29]

A DSC calorimeter (Linkam DSC 600) was used to determine the high temperature Curie transition at ~600 K. The magnetoresistance (MR) measurements were made at room temperature in an electromagnet with 1.5 T using the ac current source and a lock-in amplifier. Magnetostriction measurements were made at room temperature using a strain gauge.[24]

**Basal Pane Magnetic Isotherms:**

The fundamental property of interest in characterizing a magnetic phase at a given temperature is its spontaneous magnetization, $\sigma_{o,T}$ - the magnetization within a single magnetic domain of the crystal at zero-field. Practically, however, the measurement of $\sigma_{o,T}$ is complicated by the fact that a magnetic material generally exists as an aggregate of magnetic domains and its direct measurement within a single domain is not feasible. While the application of magnetic field can erase the domains, the resulting quantity is not spontaneous magnetization $\sigma_{o,T}$ but the specific magnetization at a given field $\sigma_{H,T}$. To overcome this problem, three principal methods have historically been developed (the so-called magnetic isotherm method, method of curves of constant magnetization, and measurements based on magneto-caloric effect).[30, 31] The magnetic isotherm method (used in the present study) involves the measurement of magnetization versus field, followed by extrapolation of the linear, high-field portion of the curve to zero-field.[32]

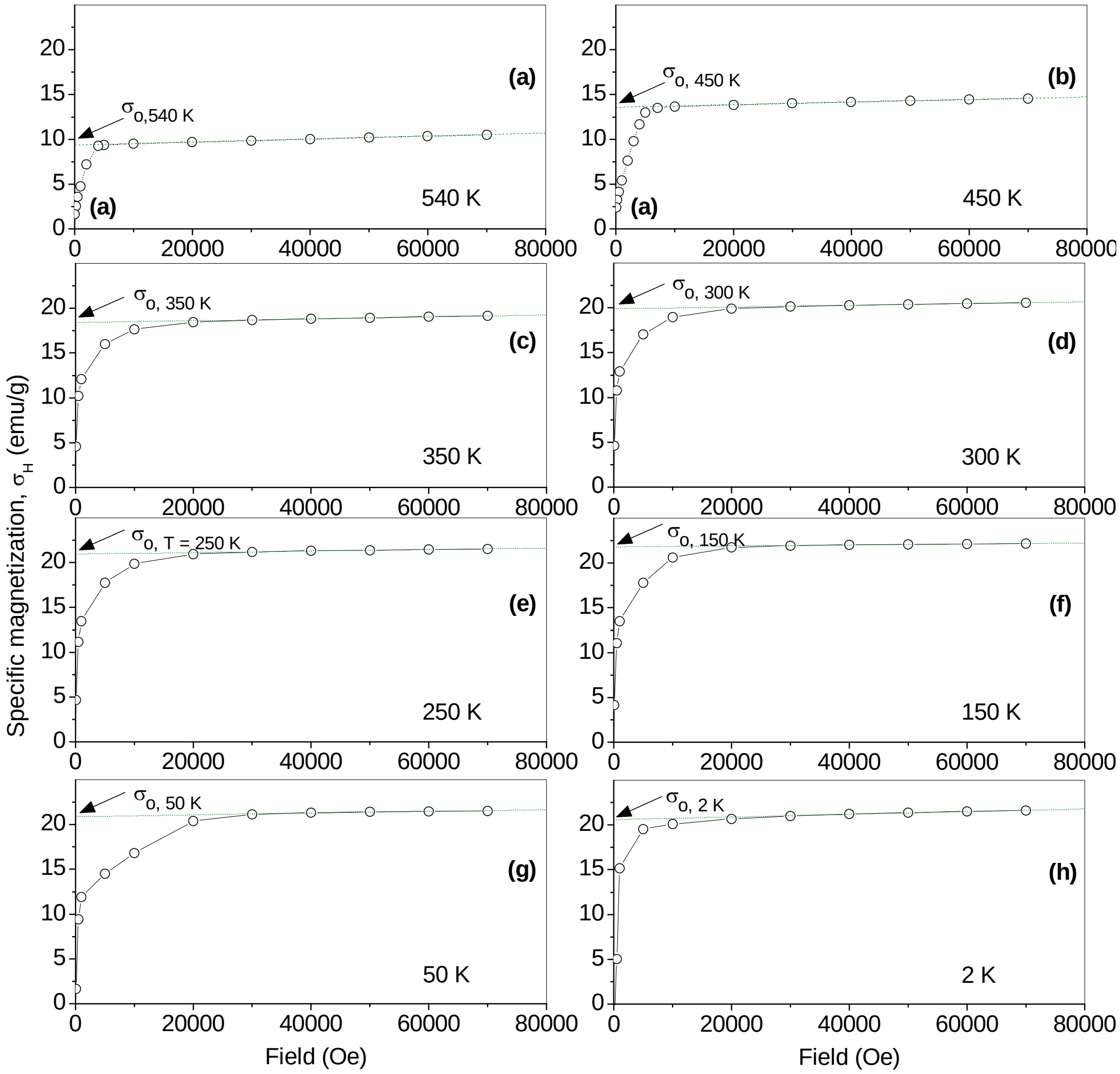

**Figure S3(a-g)**. Examples of in-plane magnetization curves at selected temperatures from which $\sigma_{o,T}$ is extracted. Extrapolation of spontaneous magnetization ($\sigma_{o,T}$) is done by the magnetic isotherm method. Measurements were made in PPMS by cooling the sample.

Measurements are made at various temperatures, from which the temperature dependence of $\sigma_{o,T}$ can be determined. Figure S3(a-g) show examples of magnetization curves with applied field within the basal plane at various temperatures. Within the basal plane the direction of applied field was along the prism plane normal. Measurements were made by first heating the sample well above the Curie temperature of 603 K, followed by measurement of the magnetization curves at various temperatures. Notice that the spontaneous magnetization doubles itself when the temperature is reduced from 540 K to 150 K. Below ~150 K, $\sigma_{o,T}$ begins to decrease. From such curves the temperature dependence of spontaneous magnetization can be determined, as shown in Fig. 1(a) of the main text.

## Supplementary References